\documentclass[pra,aps,twocolumn,showpacs]{revtex4}
\usepackage{amssymb,float}
\usepackage{amsmath}
\usepackage{amsfonts}
\usepackage[dvipdfmx]{graphicx}   
\usepackage{multirow}
\usepackage[section]{placeins}
\usepackage{ulem}
\usepackage{dcolumn}
\usepackage{bm}    
\usepackage{mediabb}

\usepackage{color}

\newcommand{\bolT}{\text{\bf T}}

\newcommand{\boll}{\mathbf{l}}
\newcommand{\bolm}{\mathbf{m}}
\newcommand{\bolx}{\mathbf{x}}
\newcommand{\boly}{\mathbf{y}}
\newcommand{\bolu}{\mathbf{u}}
\newcommand{\bolv}{\mathbf{v}}

\newcommand{\boldv}{\mathbf{d}}

\newcommand{\bra}[1]{\langle #1 |}  
\newcommand{\ket}[1]{| #1 \rangle}  
\newcommand{\VEV}[1]{\langle #1 \rangle}

\newsavebox{\dotdot}
\savebox{\dotdot}[3mm]{\shortstack{\circle*{0.8}\\ \\ \circle*{0.8}}}

\begin{document}

\title{Quantum solitons with emergent interactions 
in a model of cold atoms on the triangular lattice}

\author{Hiroaki T. Ueda$^{1,3}$\thanks{E-mail address: hueda@pu-toyama.ac.jp}, 
Yutaka Akagi$^{2,3}$, and Nic Shannon$^3$}

\affiliation{
$^1$Faculty of Engineering, Toyama Prefectural University, Izumi 939-0398, Japan\\
$^2$Department of Physics, Graduate School of Science, The University of Tokyo, Hongo, Tokyo 113-0033, Japan\\
$^3$Okinawa Institute of Science and Technology Graduate University, Onna-son, Okinawa 904-0395, Japan}

\begin{abstract}

Cold atoms bring new opportunities to study quantum magnetism, and 
in particular, to simulate quantum magnets with symmetry greater than $SU(2)$.  
Here we explore the topological excitations which arise in a model of 
cold atoms on the triangular lattice with $SU(3)$ symmetry. 
Using a combination of homotopy analysis and analytic field--theory  
we identify a new family of solitonic wave functions characterised by integer charge 
\mbox{${\bf Q} = (Q_A, Q_B, Q_C)$}, with \mbox{$Q_A + Q_B + Q_C = 0$}.
We use a numerical approach, based on a variational wave function, to 
explore the stability of these solitons on a finite lattice.
We find that solitons with charge $\mbox{${\bf Q} = (1,1,-2)$}$ spontaneously decay into 
a pair of solitons with elementary topological charge, and emergent interactions.
This result suggests that it could be possible to realise a new
class of interacting soliton, with no classical analogue, using cold atoms.
It also suggests the possibility of a new form of quantum spin liquid, 
with gauge--group U(1)$\times$U(1).
%
\end{abstract}


\maketitle


While many aspects of quantum systems can be understood 
at a local level, it is their non--local, topological properties
which offer the deepest and most surprising insights.
Topology underlies our understanding of such highly correlated
systems as $^3$He and the fractional quantum Hall effect \cite{thouless98}, 
and has come to play an important role in the theory of 
metals \cite{nagaosa10,xu15}, superconductors \cite{qi11}, 
and even systems as seemingly conventional as band insulators 
\cite{hasan10}. 
The study of topological excitations in magnets has also 
enjoyed a recent renaissance, as it has become possible
to study the interplay between topological excitations, 
such as skyrmions, and itinerant electrons \cite{nagaosa13}.


At the same time, cold atoms have brought an opportunity to study quantum 
many-body physics in a new context \cite{jaksch98,bloch08}.	
The phases realized include analogues of both magnetic metals 
and magnetic insulators \cite{joerdens08,schneider08}, with the exciting 
new possibility of extending spin symmetry from the familiar SU(2) to SU(N) 
\cite{wu03,honerkamp04,gorelik09,cazalilla09,gorshkov10,taie12,bonnes12}.
Spin models with enlarged symmetry bring with them the possibility 
to study new kinds of topological excitation, but to date, these remain 
relatively unexplored \cite{cazalilla09}.


In this Communication, we consider the topological excitations which arise in an SU(3) 
antiferromagnet that could be realised quite naturally using cold atoms.  
Starting from the most general model for SU(3) spins on the triangular lattice, 
we use a combination of field--theory and homotopy analysis to categorise 
stable topological defects.
We find a new kind of stable lump soliton with 2$^{nd}$ homotopy group
\mbox{$\pi_2  = {\mathbb Z} \times {\mathbb Z}$}, characterized by the integer charges 
$(Q_A,Q_B,Q_C)$, with $Q_A+Q_B+Q_C=0$, and obtain an analytic wave function 
for solitons with charge $(-Q,Q,0)$.   
We then study these solitons numerically, introducing a new quantum 
variational approach. 
The numerical analysis confirms the stability of the soliton with elementary 
topological charge $(-1,1,0)$.
We find that the soliton for higher topological charge with $(1,1,-2)$ 
decompose into solitons with elementary charge, with emergent repulsive 
interactions.
An example of a soliton with elementary charge ${\bf Q} = (-1,1,0)$ is 
shown in Fig.~\ref{fig:1}


\begin{figure}[t]
\includegraphics[width=1.0\columnwidth]{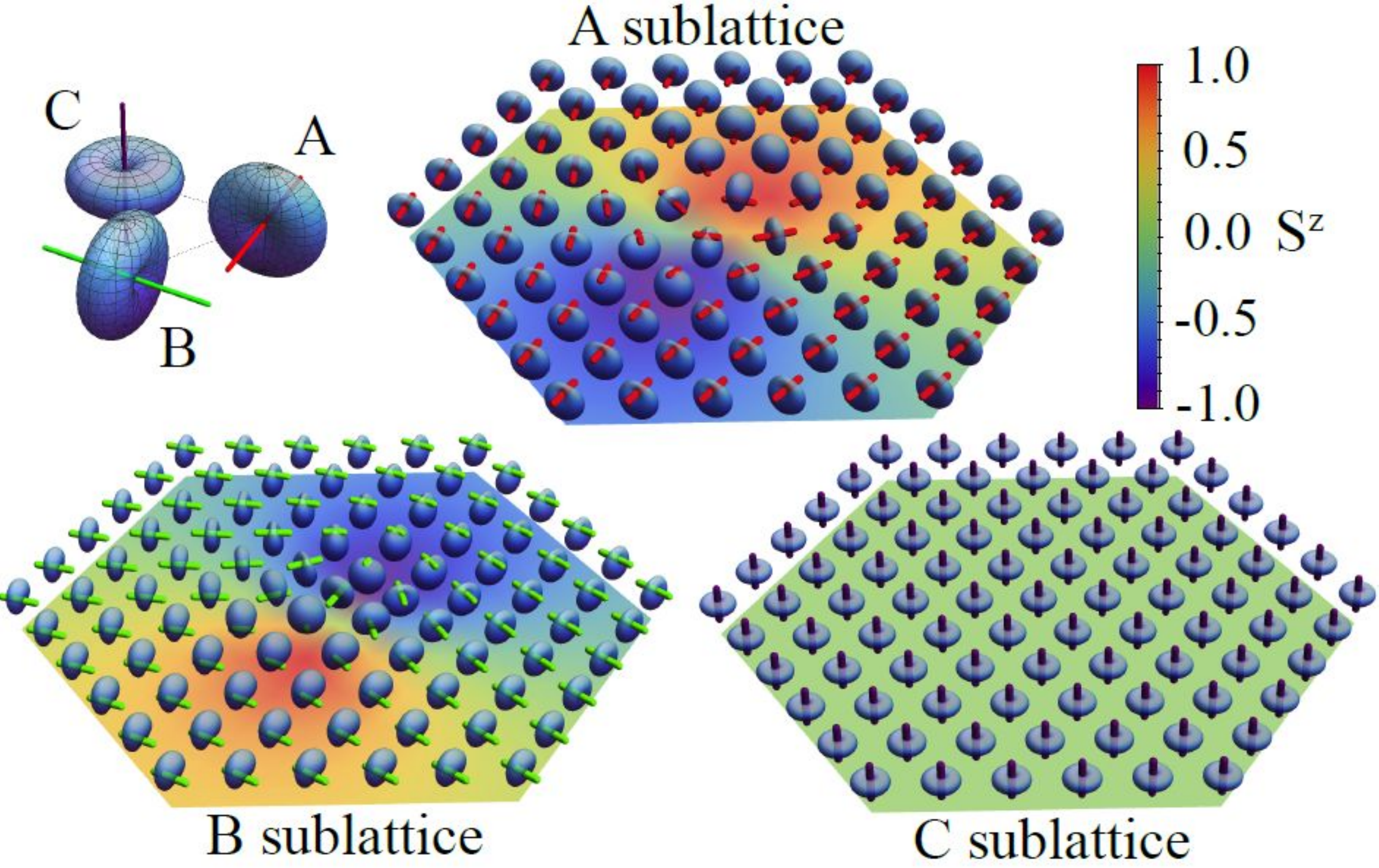}
\caption{(Color online).  
Illustration of the spin configuration of a soliton of elementary charge 
${\bf Q} = (-1,1,0)$ on the triangular lattice, decomposed into each of the 
three sublattices.
The new soliton is composed of orthogonal CP$^2$ solitons on the A and B 
sublattices, while the C sublattice remains topologically trivial.  
The probability--surface for each spin-1 moment (defined in the supplemental materials), 
is rendered in blue, while the color underlay shows the dipole moment, $S^z$, 
induced by the soliton.
Results are taken from the exact, analytic, wave function 
Eq.~(\ref{eq:analytic-wave-function}).
The reference state $\ket{r}$, Eq.~(\ref{eq:reference-state}), 
is shown as an inset. 
} 
\label{fig:1}
\end{figure}


The model we consider is the SU(3)--symmetric generalisation of the 
Heisenberg model on a triangular lattice
\begin{eqnarray}
\mathcal{H}^{\sf exchange}_{\text{SU(3)}} 
	    = J \sum_{\VEV{\boll,\bolm}} {\mathcal P}^{\text{SU(3)}}_{\boll,\bolm}
	\label{eq:HeisenbergSU3}
\end{eqnarray}
where ${\mathcal P}^{\text{SU(3)}}_{\boll,\bolm}$ is a
permutation operator exchanging the states of the atoms on sites $\boll$ and $\bolm$. 
Magnetism of this type can be realised using the hyperfine multiplets of 
repulsively--interacting Fermi atoms \cite{honerkamp04,gorelik09,bauer12}, 
an approach which has already been shown to work for SU(2) Mott Insulators 
\cite{joerdens08,schneider08}, and was recently also demonstrated 
in experiment for an SU(6) Mott insulator \cite{taie12} --- indeed SU(N) systems  
with $N > 2$ may have advantages for cooling \cite{taie12,bonnes12}.


We consider the fundamental representation of SU(3), for which
\begin{eqnarray}
	  \mathcal{H}^{\sf exchange}_{\sf SU(3)} 
	        = J \sum_{\VEV{\boll,\bolm}} 
	                 \bolT_\boll \cdot \bolT_\bolm \; ,
\end{eqnarray}
where $\bolT$ is an 8--component vector, comprising the 8 independent 
generators of SU(3) \cite{peskin95}.
These generators can be expressed terms of a quantum spin--1 
\cite{papanicolaou88,batista02,batista04,penc11-book.chapter,smerald13-PRB88}, with
\begin{eqnarray}
(T^1,T^2,T^3) = (S^x,S^y,S^z) \; ,
\end{eqnarray}
while the remaining five components of $\bolT$ are given by the quadrupole moments 
%
\begin{eqnarray}
	   \left(
	   \begin{array}{c}
	   T^4 \\
	   T^5 \\
	   T^6 \\
	   T^7 \\
	   T^8
	   \end{array}
	   \right) 
	=
	    \left(
	   \begin{array}{c}
	   (S^{\sf x})^2 - (S^{\sf y})^2 \\
	   \frac{1}{\sqrt{3}}[2(S^{\sf z})^2-(S^{\sf x})^2 - (S^{\sf y})^2] \\
	   S^{\sf x}S^{\sf y}+S^{\sf y}S^{\sf x} \\
	   S^{\sf y}S^{\sf z}+S^{\sf z}S^{\sf y} \\
	   S^{\sf x}S^{\sf z}+S^{\sf z}S^{\sf x}
	   \end{array}
	   \right) \; ,
\end{eqnarray}
familiar from the theory of liquid crystals \cite{degennes95}.
Viewed this way, the SU(3) symmetry takes on a clear physical meaning --- 
spin quadrupoles and dipoles enter into Eq.~(\ref{eq:HeisenbergSU3}) 
on an equal footing.
In addition, SU(3) rotations permit dipole moments to be transformed 
continuously into quadrupole moments, and vise--versa 
\cite{batista02,batista04,penc11-book.chapter,smerald13-PRB88}.
This is a process which has no analogue in classical magnets 
or liquid crystals, and has vital implications for topological defects.


In addition to providing a natural description of an SU(3)--symmetric Mott insulator, 
Eq.~(\ref{eq:HeisenbergSU3}) can also be thought of as a SU(3)--symmetric limit 
of the spin--1 bilinear biquadratic (BBQ) model \cite{papanicolaou88}, 
a model which may also be realised using cold atoms \cite{imambekov03,forgesdeparny14-PRL113}. 
The \mbox{spin--1} BBQ model has been widely studied on the triangular lattice, 
where it supports both conventional magnetic ground states, and quadrupolar phases analogous
to liquid crystals 
\cite{tsunetsugu06, bhattacharjee06, laeuchli06, tsunetsugu07, stoudenmire09,penc11-book.chapter, grover11, kaul12, smerald13-PRB88, smerald13-book,voell15}.
These, in turn, play host to a rich variety of topological excitations 
\cite{belavin75,kawamura84,ivanov03,ivanov07,grover11,xu12,galkina15,yutaka-unpub}. 
In particular, unconventional solitons have been shown to arise in 
the SU(3)--symmetric Heisenberg ferromagnet, 
i.e. Eq.~(\ref{eq:HeisenbergSU3}), for $J < 0$ \cite{ivanov08}.   
As yet, however, little is known about the topological defects of Eq.~(\ref{eq:HeisenbergSU3}) 
for {\it antiferromagnetic} interactions $J>0$, the case which arises 
most naturally for cold atoms \cite{honerkamp04,gorelik09,bauer12}.


The topological excitations of a given state follow from the structure of its 
ground--state manifold \cite{mermin79}.	
The ground state of Eq.~(\ref{eq:HeisenbergSU3}) on the triangular lattice, for $J>0$,
is known to break spin--rotation symmetry, and to have 
3--sublattice order \cite{laeuchli06,penc11-book.chapter,bauer12}.
However, since the SU(3) symmetry permits rotations between quadrupole and dipole 
moments of spin, this ordered state does not correspond to any single 3--sublattice 
dipolar or quadrupolar state, but rather a continuously connected manifold 
\cite{smerald13-PRB88,smerald13-book}.
In what follows we construct a representation of this ground state manifold 
and use it to classify the topological excitations of Eq.~(\ref{eq:HeisenbergSU3}).
While these results are completely general, they can most easily be understood 
through a mean--field description of Eq.~(\ref{eq:HeisenbergSU3}).   


Following \cite{laeuchli06,smerald13-PRB88}, we write 
\begin{equation}
\mathcal{H}^{\sf MFT}_{\text{SU(3)}}
	= 2J \sum_{\VEV{\boll,\bolm}} |\boldv_\boll\cdot\bar{\boldv}_\bolm|^2
	+ \text{const} \; ,
	\label{eq:H.MFT}
\end{equation}
where $\boldv$ is a complex vector with unit norm, expressing
the most general wave function for a quantum spin--1 
\begin{equation}
	\ket{\boldv} = d^x \ket{x} + d^y \ket{y} + d^z \ket{z}\ ,
\end{equation}
in terms of a basis of orthogonal spin quadrupoles
\begin{equation}
	\ket{x} \! = \! i \frac{\ket{1} \! - \! \ket{-1}}{\sqrt{2}}\ ,\ 
	\ket{y} \! = \! \frac{\ket{1} \! + \! \ket{-1}}{\sqrt{2}}\ ,\ 
	\ket{z} \! = \! -i\ket{0}\ ,
\label{ketxyz}
\end{equation}
and $\ket{1}$ is the state with $S^z = 1$, etc.  
For $J > 0$, Eq.~(\ref{eq:H.MFT}) supports a manifold of 3--sublattice ground states 
satisfying the orthogonality condition
\begin{equation}
\boldv_\lambda \cdot \bar{\boldv}_{\lambda'} = \delta_{\lambda \lambda'} 
\label{eq:orthogonality.condition}
\end{equation}
where $\lambda , \lambda' = \text{\{A, B, C\}}$.   
The simplest wave function satisfying Eq.~(\ref{eq:orthogonality.condition})
is the 3--sublattice antiferroquadrupolar (AFQ)  state 
\begin{equation}
\boldv_{A} = (1,0,0) \; , \; 
\boldv_{B} = (0,1,0) \; , \; 
\boldv_{C} = (0,0,1) \; ,
\label{eq:reference-state}
\end{equation}
illustrated in the inset of Fig.~\ref{fig:1}.   
We take this a reference state, 
and denote it $\ket{r}$.


We are now in a position to determine the symmetry of the ground state manifold,
and the topological excitations which follow from it.
The universal covering group, G \cite{mermin79}, is given by a global SU(3) 
rotation acting on $\ket{r}$.
However the order parameters $\VEV{\bolT_A}$, $\VEV{\bolT_B}$, $\VEV{\bolT_C}$
are unchanged if the following matrices act on $\ket{r}$:
\begin{equation}
f=\left( \begin{array}{ccc}
e^{i\theta_1} & 0 & 0 \\
0 & e^{-i\theta_1} & 0 \\
0 & 0 & 1
\end{array} \right)\times
\left( \begin{array}{ccc}
e^{i\theta_2} & 0 & 0 \\
0 & 1 & 0 \\
0 & 0 & e^{-i\theta_2}
\end{array} \right)\ ,\ 
\label{isotropy1}
\end{equation} 
where $\theta_1$ and $\theta_2$ are two freely--chosen phases.
It follows that the isotropy subgroup of our model is 
\begin{eqnarray}
H = U(1) \times U(1) \; ,
\end{eqnarray}
where each U(1) refers to a phase $0 \leq \theta < 2\pi$.    
Hence, the ground state manifold 
of Eq.~(\ref{eq:HeisenbergSU3}), for $J > 0$, is given by 
$\text{G/H} = \text{SU(3)} /( \text{U(1)} \times \text{U(1)} )$.  
The topological excitations supported by Eq.~(\ref{eq:HeisenbergSU3})
follow directly from this result, and a standard application of homotopy 
theory \cite{mermin79} gives
\begin{align}
&\pi_1( \text{SU(3)}/(\text{U(1)} \times \text{U(1)} ) ) = 0 \; , \\
&\pi_2( \text{SU(3)}/(\text{U(1)} \times \text{U(1)} ) ) = \mathbb{Z} \times \mathbb{Z} \; ,
\label{eq:homotopy_AFQ_SU3}
\end{align}
where the trivial first homotopy group $\pi_1$ implies the absence of 
point--like defects, while the non--zero second homotopy group 
$\pi_2$ implies the existence of a stable soliton classified by two integers.
This result should be contrasted 
with that obtained for {\it ferromagnetic} interactions, $J < 0$, in which case
\mbox{$\text{G/H} = \text{SU(3)}/(\text{SU(2)} \times \text{U(1)}) = \text{CP}^2 $}, 
and stable solitons are classified by a single integer $\mathbb{Z}$ \cite{dadda78,ivanov08}. 


We can gain more understanding of this new class of solitons by constructing
explicit wave function describing them.
To this end, we consider the continuum limit of Eq.~(\ref{eq:H.MFT}) 
and, following Refs.~\cite{smerald13-PRB88,smerald13-book}, write 
\begin{eqnarray}
\mathcal{H}^{\sf eff}_{\text{SU(3)}} 
	&=& \frac{2J}{\sqrt{3}} \int d{\bf x} \sum_{\mu = x, y}
	\left( | \boldv^\ast_A \cdot \partial_\mu \boldv_B |^2 
	+ | \boldv^\ast_B \cdot \partial_\mu \boldv_C |^2  \right. \nonumber\\
	&& \left. \qquad \qquad + \ | \boldv^\ast_C \cdot \partial_\mu \boldv_A |^2 \right) \; .
	\label{eq:Heff}
\end{eqnarray}
Building on previous work on CP$^2$ solitons \cite{dadda78}, 
we introduce a real scalar field 
\begin{equation}
A_\mu^{\lambda}
	= \frac{1}{2} i \left[ \boldv^\ast_\lambda \cdot \partial_\mu \boldv_\lambda
	- ( \partial_\mu \boldv^\ast_\lambda ) \cdot \boldv_\lambda \right] \; ,
\end{equation}
and use identities of the form
\begin{equation}
\begin{split}
& |\boldv^\ast_B \cdot \partial_\mu \boldv_A |^2 
+ |\boldv^\ast_C \cdot \partial_\mu \boldv_A |^2\\
& = | \langle \boldv_B |\partial_\mu \ket{\boldv_A} |^2
+| \langle \boldv_C |\partial_\mu \ket{\boldv_A}|^2\\
& = (\partial_\mu \bra{\boldv_A})\partial_\mu \ket{\boldv_A} 
- |\langle \boldv_A|\partial_\mu \ket{\boldv_A}|^2 \; , \end{split}
\end{equation}
where 
\mbox{$\sum_{\lambda=A,B,C} \ket{\boldv_\lambda} \bra{\boldv_\lambda} =1$},
to express Eq.~(\ref{eq:Heff}) as
\begin{eqnarray}
\label{eq:elegant.form.of.H}
\mathcal{H}^{\sf eff}_{\text{SU(3)}}
	&=& \sum_{\lambda = A,B,C} 
	\frac{J}{\sqrt{3}}
	\int d{\bf x} 
	\sum_{\mu = x, y} 
	| D^\lambda_\mu \boldv_\lambda |^2 \; , \\
         D^\lambda_\mu &=& \partial_\mu + i A^\lambda_\mu \; .
\end{eqnarray}
Here $A^\lambda_\mu$  transforms as 
\mbox{$A^\lambda_\mu \to A^\lambda_\mu - \partial_\mu \Lambda$} under 
the gauge transformation $\boldv^\prime (x) = e^{i\Lambda}\boldv (x)$.
We note that the orthogonality condition, Eq.~(\ref{eq:orthogonality.condition}), 
implies that any two of the vectors $\boldv_\lambda$ uniquely determine 
the third, leaving only phase degrees of freedom.


Solitonic solutions of Eq.~(\ref{eq:elegant.form.of.H}) are characterised by a finite 
topological charge.
In order to parameterise this, we consider the based, second homotopy group 
\begin{eqnarray}
\pi_2(SU(3)/(U(1)\times U(1)) \; , \; b)
\nonumber
\end{eqnarray}
where the base--point $b$ is given by 
\begin{eqnarray}
b = f \ket{r} \; .
\end{eqnarray}
Following Ref.~\cite{mermin79}, this is determined by the {\it first} homotopy 
group of the isotropy subgroup $\pi_1(H)  = {\mathbb Z} \times {\mathbb Z}$
[cf. Eq.~(\ref{eq:homotopy_AFQ_SU3})].  
The two independent integers $\mathbb Z$ distinguish the different solitons 
which are possible within this order--parameter space and, thereby, their 
topological charge.


We can evaluate the topological charge associated with a given soliton
by considering how the vectors $\boldv_\lambda$ evolve on a closed path $C(l)$ 
for $0\leq l \leq 1$ (with $C(0)=C(1)$), 
which encloses the soliton in the two--dimensional lattice space.  
For practical purposes, this closed path could be the boundary of a finite--size cluster.
The simplest example is a soliton characterised by the integers 
\mbox{${\mathbb Z} \times {\mathbb Z} = (1,0)$}, 
in which case the state on the path $C(l)$ is given by
\begin{equation}
f_0(l)\ket{r}=
\left( \begin{array}{ccc}
e^{-i2\pi l} & 0 & 0 \\
0 & e^{i2\pi l} & 0 \\
0 & 0 & 1
\end{array} \right)\ket{r}\ ,\ 
\label{eq:loopf1}
\end{equation}
where $0\leq l \leq 1$ and $C(0)=C(1)$.  
The contribution to the topological charge from each sublattice can be identified
with the winding number associated with the diagonal elements of $f_0$.
In the case of the A--sublattice 
\begin{equation}
\boldv_A (l) = f_0(l) \boldv_A = (e^{-i2\pi l},0,0) \; ,
\end{equation}
and it follows that the winding number on the path $C(l)$ is unity, 
and the associated topological charge is given by $Q_A = 1$.
More formally, we can calculate this winding number as
\begin{eqnarray}
&& Q_A = \frac{i}{2\pi} \oint_{C(l)} d{\mathbf l} \cdot 
\left[ \boldv^\ast_A(\bolx) {\bf \nabla} \boldv_A (\bolx) \right] \nonumber\\
&& = \frac{i}{2\pi} \int d{\bf x}\ (\partial_x \boldv^\ast_A(\bolx))(\partial_y \boldv_A(\bolx))
- (\partial_y \boldv^\ast_A(\bolx))(\partial_x \boldv_A(\bolx)) \nonumber\\
&& = \frac{i}{2\pi} \int d{\bf x}\  
	\epsilon_{\mu\nu}\ (D_\mu^A \boldv_A)^\ast 
	\cdot D^A_\nu \boldv_A   = 1 \; ,
\end{eqnarray}
where the two--dimensional integral $\int d{\bf x}$ is carried 
out over the area enclosed by the path $C(l)$.
By inspection of Eq.~(\ref{eq:loopf1}), the contribution of the B--sublattice is
$Q_B = - Q_A  = -1$, while the contribution from the (topologically--trivial) 
\mbox{C--sublattice} vanishes. 
Therefore, for this example, \mbox{${\mathbf Q} = (Q_A,Q_B,Q_C)=(1,-1,0)$}.  


This approach to evaluating the topological charge remains valid, 
regardless of the base point, and can be applied to any spin 
configuration.   
So quite generally, we can write 
\begin{equation}
Q_\lambda = \frac{i}{2\pi} \int d{\bf x}\  
	\epsilon_{\mu\nu}\ (D_\mu^\lambda \boldv_\lambda ) ^\ast
	\cdot D^\lambda_\nu \boldv_\lambda \; ,
\label{eq:topological.charge}
\end{equation}
subject to the constraint 
\begin{equation}
Q_A + Q_B + Q_C = 0 \; .
\end{equation}
This constraint follows from the structure of the isotropy subgroup $H$, 
i.e. the fact that there are only two undetermined angles in Eq.~(\ref{isotropy1}). 
Thus, while it is covenant to quote the topological charge as a vector ${\mathbf Q}$
with three components, there are only ever two independent degrees of freedom.
Ultimately, this reflects the orthogonality condition on the three vectors 
${\mathbf d}_{\lambda=A,B,C}$, as defined in Eq.~(8).
We note that our definition of the topological charge, Eq.~(\ref{eq:topological.charge}),  
differs by a sign convention from that definition used by in Ref.~\cite{ivanov08}.


Viewed this way, the continuum theory, Eq.~(\ref{eq:elegant.form.of.H}), comprises 
three copies of a CP$^2$ nonlinear sigma model, linked by the orthogonality condition, 
Eq.~(\ref{eq:orthogonality.condition}).
The Cauchy--Schwartz inequality \cite{dadda78} enables us to place a lower 
bound on the energy of a soliton, based on its charge
\begin{equation}
E_{\bf Q} 
	\geq \frac{2\pi J}{\sqrt{3}} \left( |Q_A|+|Q_B|+|Q_C| \right) \; .
	\label{eq:bound}
\end{equation}
Equality in Eq.~(\ref{eq:bound}), also known the as 
the Bogomol'nyi--Prasad--Sommerfield (BPS) 
bound \cite{manton04}, is achieved where 
\begin{equation}
D_\mu^\lambda \boldv_\lambda 
	= \pm\ i \epsilon_{\mu\nu}\ D_\nu^\lambda \boldv_\lambda 
	\quad \forall \quad \lambda = A,B,C \; .
	\label{eq:constraint}
\end{equation}
We have been able to construct an exact wave function satisfying the 
Eq.~(\ref{eq:constraint}), for the special case 
\mbox{${\bf Q} = (-Q, Q, 0)$}.  
This corresponds to Q pairs of orthogonal CP$^2$ solitons \cite{dadda78,ivanov08},
found on two of the three sublattices, while the third remains topologically trivial.
Specifically, we find
\begin{equation}
\begin{split}
\boldv_A (z) 
	& = \frac{\xi \bolu_1 + \left[ \Pi_{k=1}^Q (z - z_k) \right] \bolv_1 }
	{\sqrt{\xi^2 + \Pi_{k=1}^Q | z - z_k |^2)}} \; ,\\
\boldv_B (z) 
	& =\frac{-\xi \bolv_1 +  \left[ \Pi_{k=1}^Q (z^\ast - z^\ast_k) \right] \bolu_1}
	{\sqrt{(\xi^2 +   \Pi_{k=1}^Q | z^\ast - z^\ast_k |^2)}} \; ,\\
\boldv_C (z) 
	& = \boly_1\ \; .
\end{split}
\label{eq:analytic-wave-function}
\end{equation}
where $z = x+iy$ is a complex coordinate, $\bolu_1$, $\bolv_1$ and $\boly_1$ 
are complex orthonormal vectors, taken to be 
\begin{eqnarray}
\bolv_1=(1,0,0) \; , \; \bolu_1=(0,1,0) \; , \;  \boly_1=(0,0,1) \; .
\end{eqnarray}
The coordinate $z_k$ specifies the positions of each soliton.
Since the energy of this family of solitons is completely determined
by its charge [cf.~Eq.~(\ref{eq:bound})], it is independent 
of their size, which is set by the real parameter~$\xi$.
The wave function, Eq.~(\ref{eq:analytic-wave-function}), for a soliton with 
elementary charge ${\bf Q} = (-1,1,0)$, is illustrated in Fig.~\ref{fig:1}.


\begin{figure}[t]
\includegraphics[width=0.95\columnwidth]{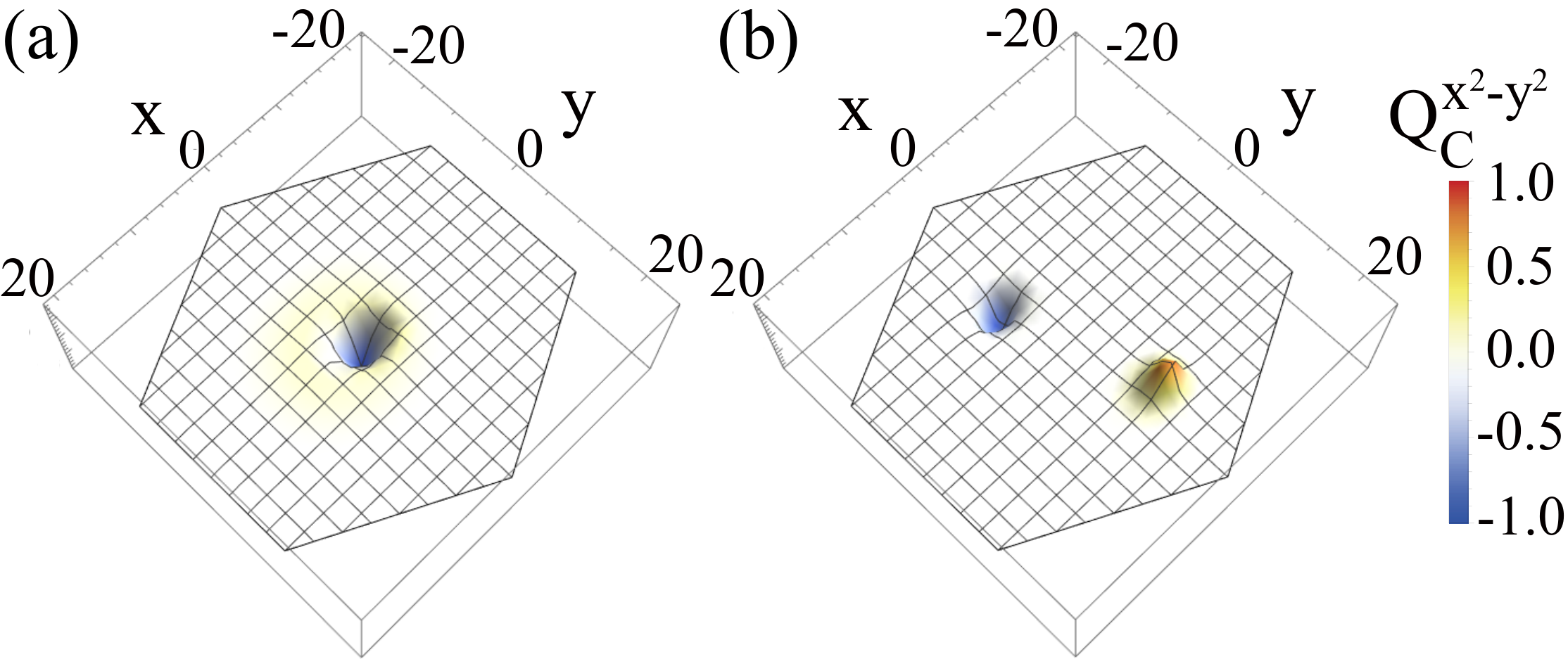}
\caption{(Color online). 
Illustration of how 
repulsive interactions cause a single soliton with charge
${\bf Q} = (1,1,-2)$ to break into two separate pieces.
(a) Quadrupole moment $Q^{x^2-y^2}_C$ on the C sublattice 
associated with the trial wave function Eq.~(\ref{eq:trial_wave_function}).
(b)~Quadrupole moment after numerical minimisation of a variational wave function.  
} 
\label{fig:2}
\end{figure}


\begin{figure*}[t]
\includegraphics[width=1.8\columnwidth]{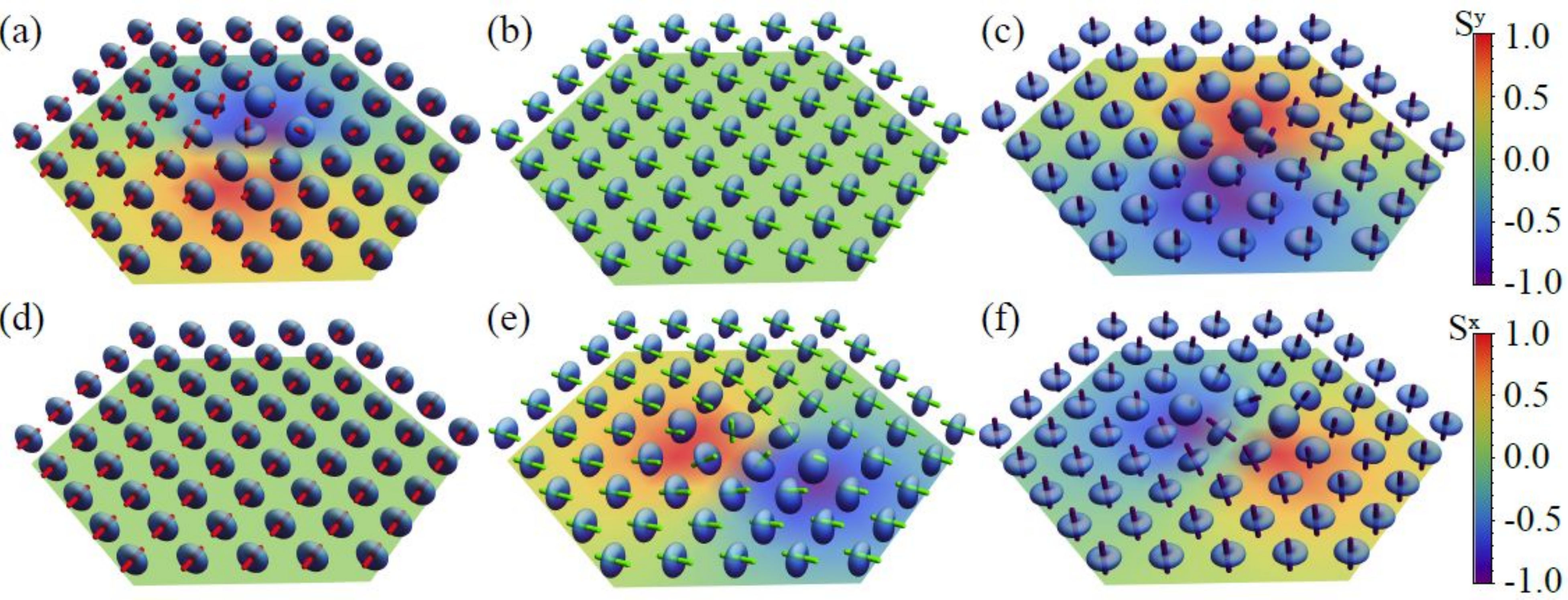}
\caption{(Color online).
Details of the variational wave function describing a state with topological charge 
${\bf Q} = (1, 1, -2)$, resolved onto A, B and C sublattices.
(a)--(c) wave function near to the maximum in Fig.~\ref{fig:2}(b),  
showing a soliton with elementary charge ${\bf Q} = (1, 0, -1)$.
(d)--(f) wave function near to the minimum in Fig.~\ref{fig:2}(b),  
showing a soliton with elementary charge ${\bf Q} = (0, 1, -1)$.
The probability--surface for each spin-1 (defined in the supplemental 
materials), is rendered in blue, while the color underlay shows the dipole 
moments $S^y$ [(a)--(c)], and $S^x$ [(d)--(f)], induced by the soliton.
} 
\label{fig:3}
\end{figure*}


In order to gain more insight into solitons with general charge, 
for which no closed--form solution exists, we now switch to numerical analysis.
We adopt a variational approach, based on a general product wave function, and minimise 
the energy of this wave function using simulated annealing \cite{kirkpatrick83}, 
as described in the supplemental materials.
Simulations were carried out for hexagonal clusters, 
and seeded with a trial wave function with definite topological charge ${\bf Q}$.  


In the simplest case, a soliton of elementary charge ${\bf Q} = (-1,1,0)$,
we can use the BPS bound Eq.~(\ref{eq:analytic-wave-function}) as a trial 
wave function, and simulations converge on a state like that shown 
in Fig.~\ref{fig:1}, confirming the stability of the analytic solution 
on a finite lattice \cite{supplemental}.  
We next consider the soliton with charge \mbox{${\bf Q} = (1,1,-2)$}.   
A suitable trial wave function is
\begin{align}
\begin{split}
&\boldv_C(z) = \frac{\bolu_1 + c_1 z \bolv_1 + c_2 z^2 \boly_1}
	{\sqrt{1 + |c_1 z|^2 + |c_2 z|^2}}\ ,\\
&\boldv_B(z) = \frac{- c_1 z^\ast \bolu_1 + \bolv_1}
	{\sqrt{1 + |c_1 z|^2}}\ ,\\
&\boldv_A(z) \perp \boldv_B(z)\ ,\ \boldv_C(z)\ ,
	\label{eq:trial_wave_function}
\end{split}
\end{align}
where we impose the boundary conditions 
\mbox{$\boldv_A= e^{-i \theta_z} \bolv_1$},
\mbox{$\boldv_B = -e^{-i \theta_z} \bolu_1$},
\mbox{$\boldv_C = e^{i 2 \theta_z}  \boly_1$},
at the edges of the cluster, with $\theta_z = \arg (z)$.
The coefficients $c_{1,2}$ are chosen 
so as to create a single soliton, localized at the origin,
as illustrated in Fig.~\ref{fig:2}(a).
This state has a  higher energy than the BPS bound, 
Eq.~(\ref{eq:bound}), and in numerical simulations, 
decays into two separate solitons of elementary charge, 
${\bf Q} = (0,1,-1)$ and ${\bf Q} = (1,0,-1)$, 
as shown in Fig.~\ref{fig:2}(b) and Fig.~\ref{fig:3}.   
In an infinitely extended system, these elementary charges could 
satisfy the BPS bound by separating entirely.   
However in our simulations, solitons also interact with the boundary 
of the cluster, and so only separate to finite distance.


We interpret these results as follows.
The model, Eq.~(\ref{eq:HeisenbergSU3}) supports six different types 
of ``elementary'' soliton, with charge ${\bf Q} = \pm (0,1,-1)$, etc.  
Solitons with like charge do not interact, while solitons with unlike charge 
interact repulsively.
We have also carried out preliminary simulations for solitons with 
higher topological charge, including ${\bf Q} = (0, 2,-2)$, ${\bf Q} = (1,2,-3)$ 
and ${\bf Q} = (2,2,-4)$, which confirm that this picture holds for more general 
topological charge.
These results will be reported elsewhere.


In conclusion, we have explored the topological excitations which 
arise in an SU(3)--symmetric antiferromagnet on a triangular 
lattice.   
We find that this model supports a new class of stable solitons, 
with second homotopy group 
\mbox{$\pi_2 (\text{SU(3)}/\text{U(1)} \times \text{U(1)} )
= \mathbb{Z} \times \mathbb{Z} $}.
These solitons can be characterised by integer topological 
charge \mbox{${\bf Q} = (Q_A, Q_B, Q_C)$}, 
with \mbox{$Q_A + Q_B + Q_C=0$}. 
In the case of solitons with charge ${\bf Q} = (-Q, Q, 0)$, 
we are able to construct an exact wave function satisfying 
the BPS bound, comprising orthogonal CP$^{2}$ solitons on 
two of the three sublattices A, B, C [cf. Fig.~\ref{fig:1}].
Numerical simulations were used to confirm the stability
of these solutions, and to explore the structure of solitons 
with more general topological charge.
We find that solitons with charge  \mbox{${\bf Q} = (1,1,-2)$}  
spontaneously decay into solitons with ``elementary'' charge
\mbox{${\bf Q} = (0,1,-1)$}  and \mbox{${\bf Q} = (1,0,-1)$} 
[cf.~Fig.~\ref{fig:3}].
We infer that solitons with different elementary charge interact 
repulsively.


To the best of our knowledge, these results represent the 
first example of quantum solitons characterised by two integers,  
with emergent repulsive interactions.
The model solved has direct application to experiments on cold
atoms in an optical lattice, where quantum magnets with 
SU(3) symmetry arise quite naturally \cite{honerkamp04,gorelik09,bauer12}.  
It is also interesting to speculate that these new solitons might survive as 
dynamical excitations in a \mbox{spin--1} magnet where SU(3) symmetry 
was broken, as argued for the CP$^2$ solitons found in the 
SU(3)--symmetric Heisenberg ferromagnet \cite{ivanov08}.   
And, since topological defects determine the type of quantum spin liquid
which follows when classical order melts 
\cite{chubukov94,grover11,xu12},  
these results suggest the possibility of a new class of quantum spin liquid with 
an underlying U(1)$\times$U(1) gauge structure. 
We hope that this may help to shed light on the quantum spin--liquids 
found in two--dimensional quantum magnets which might otherwise support
3--sublattice order \cite{ishida97,lee08,cheng11, xu12prl, bieri12prb}.

{\it Acknowledgments :}   
%
The authors are pleased to acknowledge helpful conversations with 
C. D. Batista, M. Kobayashi, T. Momoi, Y. Motome, T. Nitta, 
K. Penc,  A. Smerald, K. Totsuka, and D. Yamamoto,
and are grateful to F. Mila for a careful reading 
of the manuscript.  
This work was supported by the Okinawa Institute
of Science and Technology Graduate University,
and by JSPS KAKENHI Grant No. 26800209.


\begin{widetext}
Supplemental material for~: Quantum solitons with emergent interactions 
in a model of cold atoms on the triangular lattice
\end{widetext}
\maketitle

\subsection{Pictorial representation of wave functions}

\begin{figure*}[t]
\includegraphics[width=1.95\columnwidth]{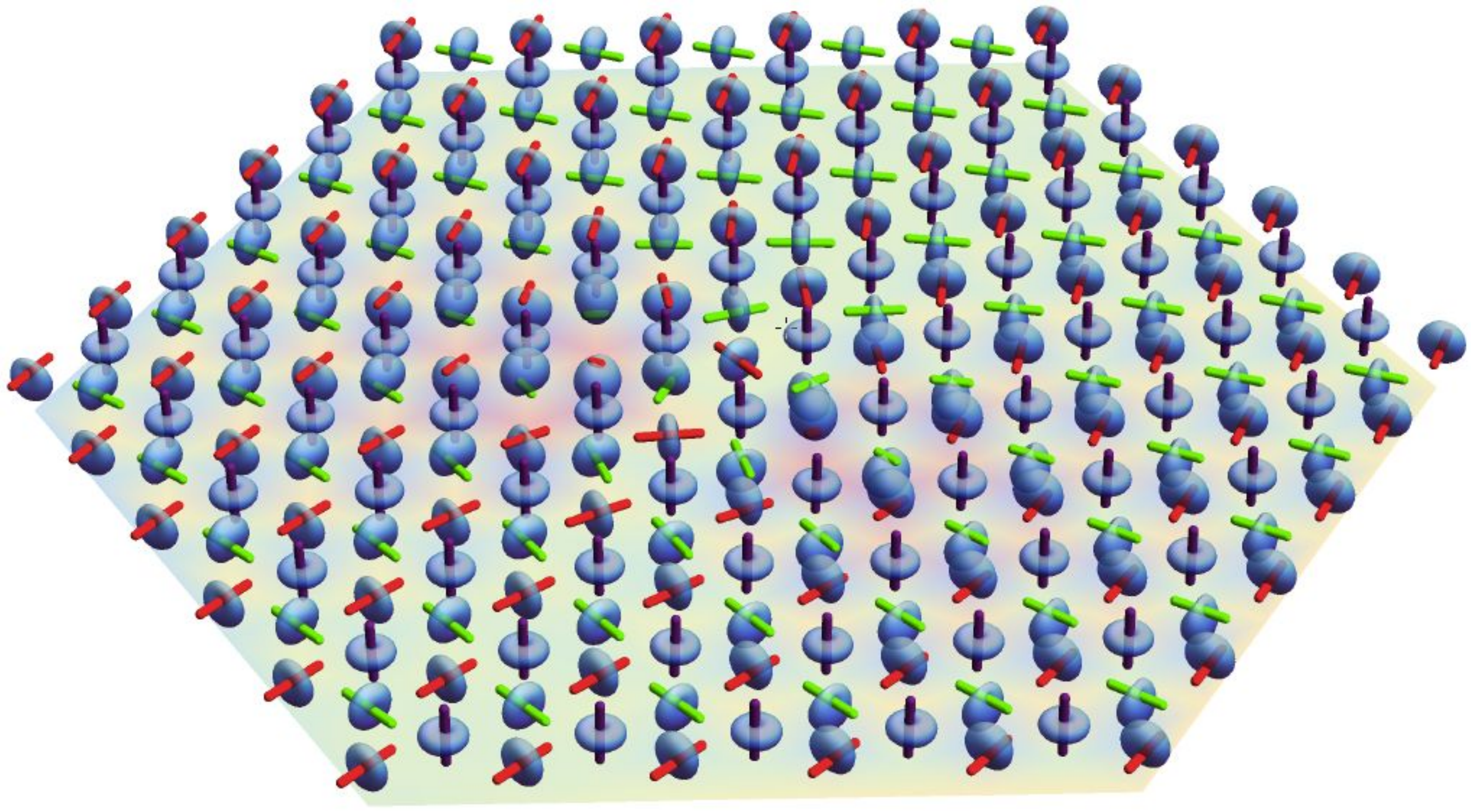}
\caption{(Color online).
Example of a soliton in an SU(3)--symmetric model of 
cold atoms on the triangular lattice.
The soliton shown has topological charge \mbox{${\bf Q} = (-1, 1, 0)$},
and is represented through the probability surface for a quantum spin-1 
at each site [cf. Eq.~(\ref{eq:prob.surface})], rendered in blue.    
Red, green and blue bars show the orientation of the quadrupole moment
on each of the three sublattices $A$, $B$, and $C$.
The color underlay shows the variation of the dipole moment, $S^z$
across the soliton.  
Results are taken from variational calculations for the SU(3)--symmetric
Heisenberg model, $\mathcal{H}^{\text{exchange}}_{\text{SU(3)}}$, 
and should be compared with Fig.~1 of the main text.  
}
\label{fig:1_suppl}
\end{figure*}

It is convenient to represent the wave function of the solitons studied in 
this Rapid Communication pictorially.  
The state of quantum spin--1 can conveniently be represented as a 
probability surface
\begin{equation}
	P(\theta,\phi) = | \bra{\boldv} \boldsymbol{\Omega} \rangle |^2 
	\label{eq:prob.surface}
\end{equation}
through its projection onto a spin coherent state 
\begin{equation}
	\ket{\boldsymbol{\Omega}} = \mathcal{R}(\theta,\phi) \ket{1}
\end{equation}
where $\ket{1}$ is the eigenstate with $S^z = +1$, and 
$\mathcal{R}(\theta,\phi)$ is an SU(2) rotation matrix.
A pictorial representation of a soliton with elementary charge ${\bf Q} = (-1,1,0)$ 
is shown in Fig.~\ref{fig:1_suppl}, below.  
The results are taken from numerical simulation, as described below, 
and in the main text.


In many cases it is also interesting to examine the quadrupole, 
or dipole, moments of spin which are induced by the soliton, relative
to the reference state $\ket{r}$ [cf. Eq.~(9) of the main text].
These can be calculated directly from the complex vector $\boldv$ 
\cite{ivanov08,penc11-book.chapter,Toth12} --- for example the 
dipole moment associated with $\ket{\boldv}$ is given by 
\begin{align}
 {\bf S}  = 2 {\bf u} \times {\bf v} \; ,
 \label{eq:Suv}
\end{align}
where 
\begin{align}
{\bf d} = {\bf u} + i {\bf v} \; .
\end{align}


To illustrate how this works, in Fig.~\ref{fig:1_suppl} we show the results of numerical simulations 
for a soliton with elementary charge ${\bf Q} = (-1,1,0)$.
Here, the probability surface for a quantum spin-1 at each site [cf. Eq.~(\ref{eq:prob.surface})],
has been rendered in blue, while red, green and blue bars show the orientation of the 
quadrupole moment on each of the three sublattices $A$, $B$, and $C$.
The dipole moment $S^z$, which vanishes in the reference state, 
has been represented as a color underlay.


This soliton comprises orthogonal $CP^2$ solitons on two of the three sublattices
of the triangular lattice.   
Each of these $CP^2$ solitons can be thought of as a localised ``twist'' in the phase of the 
vector {\bf d}, as described in Eq.~(27) of the main text.
This twist can be smoothly matched to ferroquadrupolar ($FQ$) order at infinity.  
However it induces a $\pi$ rotation of the associated director, accompanied by a finite 
dipole moment, in the vicinity of the soliton.


More detailed description of individual $CP^2$ solitons can be found in Ref.~\cite{ivanov08} 
and Ref.~\cite{dadda78}.

\subsection{Details of numerical simulations}

In our numerical studies we consider the most general class of product wave function for 
a quantum spin--1
\begin{equation}
	\ket{ \{ \boldv_\boll \} }  =  \prod_\boll \ket{\boldv_\boll} \; ,
\end{equation}
where the product on $\boll = 1 \ldots N$ runs over all sites of a finite--size cluster, 
and $\ket{\boldv_\boll}$ is defined through Eq.~(6) of the main text.
Since $\boldv$ has unit norm, and the physical properties of $\ket{\boldv_\boll}$ 
are invariant under a change of phase, 
\begin{equation}
\ket{\boldv_\boll} \rightarrow e^{i\theta} \ket{\boldv_\boll} \; , 
\end{equation}
a wave function of this form has $4^N$ variational parameters.
These parameters are determined by simulated annealing \cite{kirkpatrick83}, 
starting from an initial ``guess'' at the soliton wave function, with definite topological
charge ${\bf Q}$.


In this approach, simulations are carried out at a temperature $T$, which is 
decreased gradually, and at each temperature the wave function is updated 
using Markov--chain Monte Carlo sampling.
Monte Carlo updates involve the random re--orientation of the vector $\boldv$ 
for each spin in turn, following the standard Metropolis algorithm.
Temperatures are reduced following a geometrical progression 
\begin{equation}
T_{k+1} = \alpha\ T_k \quad , \quad 0 < \alpha < 1 \; ,
\end{equation}
where $T_k$ is the temperature of the $k^{th}$ of $N_{\sf annealing}$ annealing steps.
Typical values are 
\begin{eqnarray}
\begin{split}
N_{\sf annealing} & = 300 & \\
\alpha & =  0.95 & \\
T_{\sf initial} &= 0.1\ J & \\
T_{\sf final} & \simeq 2.1 \times 10^{-8}\ J &
\end{split}
\end{eqnarray}


We consider hexagonal clusters with the full symmetry of the triangular lattice, and 
impose a boundary condition at the edges of the cluster consistent with the reference
state $| r \rangle$ [Eq. (9) of main text].

In addition, to ensure that simulations are carried out at fixed topological charge, 
we impose a ``smoothness condition'' 
\begin{equation}
| \partial_\mu \boldv_\lambda |^2 \leq 1 \quad , \quad \mu = x,y \quad , \quad  \lambda =A, B, C \; .
\end{equation}


\begin{thebibliography}{101}


\bibitem{thouless98}
D. Thouless, 
\textit{Topological Quantum Numbers in Nonrelativistic Physics}, 
World Scientific, Singapore(1998).

\bibitem{nagaosa10}
N. Nagaosa, J. Sinova, S. Onoda, A. H. MacDonald, and N. P. Ong,
\textit{Anomalous Hall effect}, 
Rev. Mod. Phys. \textbf{82}, 1539 (2010).

\bibitem{xu15}
S.-Y. Xu, I. Belopolski, N. Alidoust, M. Neupane,
G. Bian, C. Zhang, R. Sankar, G. Chang,  Z. Yuan, C.-C. Lee, 
S.-M. Huang,  H. Zheng, J. Ma, D. S. Sanchez, B.K. Wang,
A. Bansil, F. Chou, P. P. Shibayev, H. Lin,  S. Jia, and M. Z. Hasan, 
\textit{Discovery of a Weyl fermion semimetal and topological Fermi arcs}, 
Science \textbf{349}, 613 (2015).

\bibitem{qi11}
X.-L. Qi and S.-C. Zhang,
\textit{Topological insulators and superconductors},
Rev. Mod. Phys. \textbf{83}, 1057 (2011).

\bibitem{hasan10}
M. Z. Hasan and C. L. Kane,
\textit{Colloquium: Topological insulators},
Rev. Mod. Phys. \textbf{82}, 3045 (2010).

\bibitem{nagaosa13}
N. Nagaosa and Y. Tokura,
\textit{Topological properties and dynamics of magnetic skyrmions},
Nature Nanotech. \textbf{8}, 899 (2013).


\bibitem{jaksch98} 
D. Jaksch, C. Bruder, J. I. Cirac, C. W. Gardiner, and P. Zoller,
{\it Cold Bosonic Atoms in Optical Lattices}, 
Phys. Rev. Lett. {\bf 81}, 3108 (1998).

\bibitem{bloch08} 
I. Bloch, J. Dalibard, and W. Zwerger,
{\it Many-body physics with ultracold gases},
Rev. Mod. Phys. {\bf 80}, 885 (2008).

\bibitem{joerdens08}
R. J\"ordens, N. Strohmaier, K. G\"unter, H. Moritz, and T. Esslinger,
{\it A Mott insulator of fermionic atoms in an optical lattice},
Nature {\bf 455}, 204 (2008). 

\bibitem{schneider08}
U. Schneider, L. Hackerm\"uller, S. Will, Th. Best, I. Bloch, 
T. A. Costi, R. W. Helmes, D. Rasch, and A. Rosch,
{\it Metallic and Insulating Phases of Repulsively Interacting 
Fermions in a 3D Optical Lattice},
Science {\bf 322}, 1520 (2008).

\bibitem{wu03} 
C. Wu, J.-P. Hu, and S.-C. Zhang, 
{\it Exact SO(5) Symmetry in the Spin-$3/2$ Fermionic System},
Phys. Rev. Lett. {\bf 91}, 186402 (2003).

\bibitem{honerkamp04}
C.~Honerkamp and W.~Hofstetter,
{\it Ultracold Fermions and the SU(N) Hubbard Model},
Phys. Rev. Lett. {\bf 92}, 170403 (2004).

\bibitem{gorelik09}
E. V. Gorelik and N. Bl\"umer,
{\it Mott transitions in ternary flavor mixtures of ultracold 
fermions on optical lattices},
Phys. Rev. A {\bf 80}, 051602(R) (2009).

\bibitem{cazalilla09}
M. A. Cazalilla, A. F. Ho, and M. Ueda, 
{\it Ultracold gases of ytterbium: ferromagnetism and Mott states 
in an SU(6) Fermi system},
New J. Phys. {\bf 11}, 103033 (2009).

\bibitem{gorshkov10} 
A. V. Gorshkov, M. Hermele, V. Gurarie, C. Xu, P. S. Julienne, 
J. Ye, P. Zoller, E. Demler, M. D. Lukin, and A. M. Rey, 
{\it Two-orbital SU(N) magnetism with ultracold alkaline-earth atoms}, 
Nature Phys. {\bf 6}, 289 (2010).

\bibitem{taie12}
S. Taie, R. Yamazaki, S. Sugawa, and Y. Takahashi, 
{\it An SU(6) Mott Insulator of an atomic Fermi gas realised by 
large--spin Pomoranchuk cooling},
Nature Phys. {\bf 8}, 825 (2015). 

\bibitem{bonnes12}
L. Bonnes, K. R. A. Hazzard, S. R. Manmana, A. M. Rey, and S. Wessel, 
{\it Adiabatic Loading of One-Dimensional SU(N) Alkaline-Earth-Atom 
Fermions in Optical Lattices}, 
Phys. Rev. Lett. {\bf 109}, 205305 (2012).


\bibitem{bauer12}
B. Bauer, P. Corboz, A. M. La\"{u}chli, L. Messio, K. Penc, 
M. Troyer, and F. Mila, 
{\it Three-sublattice order in the SU(3) Heisenberg model on the 
square and triangular lattice},
Phys. Rev. B {\bf 85}, 125116 (2012).

\bibitem{peskin95}
M.~E.~Peskin and D.~V.~Schroeder, 
{\it An Introduction to Quantum Field Theory}, 
(Westview press, 1995), Chapter~15. 

\bibitem{papanicolaou88}
N. Papanicolaou,
{\it Unusual Phases in Quantum Spin-1 Systems},
Nucl. Phys. B {\bf 305}, 367 (1988).

\bibitem{batista02}
C. D. Batista, G. Ortiz, and J. E. Gubernatis,
{\it Unveiling order behind complexity: Coexistence of ferromagnetism and Bose-Einstein condensation},
Phys. Rev. B {\bf 65}, 180402(R) (2002).

\bibitem{batista04}
C. D. Batista and G. Ortiz,
{\it Algebraic approach to interacting quantum systems},
Adv. in Phys. {\bf 53}, 1 (2004).

\bibitem{penc11-book.chapter}
K. Penc and A. Lauchli, 
{\it Introduction to frustrated magnetism}, (Springer-Verlag Berlin Heidelberg 2011), chapter~13.

\bibitem{smerald13-PRB88}
A.~Smerald and N.~Shannon,
{\it Theory of spin excitations in a quantum spin-nematic state}, 
Phys.~Rev. B {\bf 88}, 184430 (2013).


\bibitem{degennes95}
P. G. De Gennes and J. Prost,
{\it The Physics of Liquid Crystals},
(International Series of Monographs on Physics, Oxford University Press, 1995.

\bibitem{imambekov03} 
A. Imambekov, M. Lukin, and E. Demler, 
{\it Spin-exchange interactions of spin-one bosons in optical lattices:
Singlet, nematic, and dimerized phases},
Phys. Rev. A {\bf 68}, 063602 (2003). 

\bibitem{forgesdeparny14-PRL113}
L. de Forges de Parny, H. Yang, and F. Mila, 
{\it Anderson Tower of States and Nematic Order of Spin-1 Bosonic Atoms on a 2D Lattice}, 
Phys. Rev. Lett. {\bf 113}, 200402 (2014).

\bibitem{tsunetsugu06}
H.~Tsunetsugu and M.~Arikawa, 
{\it Spin Nematic Phase in S=1 Triangular Antiferromagnets},
J.~Phys.~Soc.~Jpn., {\bf 75}, 083701 (2006).

\bibitem{laeuchli06}
A. L\"auchli, F. Mila, and K. Penc, 
{\it Quadrupolar Phases of the S=1 Bilinear-Biquadratic Heisenberg Model on the Triangular Lattice},
Phys. Rev. Lett. {\bf 97}, 087205 (2006); 
erratum {\it ibid}, 229901 (2006).

\bibitem{bhattacharjee06}
S. Bhattacharjee, V. B. Shenoy, and T. Senthil, 
{\it Possible ferro-spin nematic order in NiGa$_2$S$_4$},
Phys. Rev. B {\bf 74}, 092406 (2006).

\bibitem{tsunetsugu07}
H. Tsunetsugu and M. Arikawa, 
{\it The spin nematic state in triangular antiferromagnets}, 
J. Phys.: Condens. Matter {\bf 19}, 145248 (2007).

\bibitem{stoudenmire09}
E. M. Stoudenmire, S. Trebst, and L. Balents, 
{\it Quadrupolar correlations and spin freezing 
in S=1 triangular lattice antiferromagnets}, 
Phys. Rev. B {\bf 79}, 214436 (2009).

\bibitem{grover11}
T. Grover and T. Senthil,
{\it Non-Abelian Spin Liquid in a Spin-One Quantum Magnet},
Phys. Rev. Lett. {\bf 107}, 077203 (2011).

\bibitem{kaul12}
R. K. Kaul, 
{\it Spin nematic ground state of the triangular lattice $S = 1$ biquadratic model}, 
Phys. Rev. B {\bf 86}, 104411 (2012).

\bibitem{smerald13-book}
A.~Smerald,
{\it Theory of the Nuclear Magnetic 1/T1 Relaxation Rate in Conventional and 
Unconventional Magnets (Springer Theses)} 
Springer 2013.

\bibitem{voell15}
A. V\"oll and S. Wessel,
{\it Spin dynamics of the bilinear-biquadratic S=1 Heisenberg 
model on the triangular lattice: A quantum Monte Carlo study}, 
Phys. Rev. B {\bf 91}, 165128 (2015).

\bibitem{belavin75}
A. A. Belavin and A. M. Polyakov, 
{\it Metastable states of two--dimensional isotropic ferromagnets}, 
JETP Lett. {\bf 22}, 245 (1975).

\bibitem{kawamura84}
H. Kawamura and S. Miyashita, 
{\it Phase Transition of the Two-Dimensional Heisneberg 
Antiferromagnet on the Triangular Lattice}, 
J. Phys. Soc. Jpn. {\bf 53}, 4138 (1984).

\bibitem{ivanov03}
B. A. Ivanov and A. K. Kolezhuk,
{\it Effective field theory for the S=1 quantum nematic},
Phys. Rev. B {\bf 68}, 052401 (2003).

\bibitem{ivanov07}
B. A. Ivanov and R. S. Khymyn, 
{\it Soliton Dynamics in a Spin Nematic},
JETP {\bf 104}, 307 (2007).

\bibitem{xu12}
C. Xu and A. W. W. Ludwig, 
{\it Topological Quantum Liquids with Quaternion Non-Abelian Statistics}, 
Phys. Rev. Lett. {\bf 108}, 047202 (2012).

\bibitem{galkina15}
E. G. Galkina, B. A. Ivanov, O. A. Kosmachev, and Yu. A. Fridman, 
{\it Two-dimensional solitons in spin nematic states for magnets 
with an isotropic exchange interaction}, 
Low Temp. Phys. {\bf 41}, 382 (2015).

\bibitem{yutaka-unpub}
Y. Akagi, H. T. Ueda, and N. Shannon, 
{\it in preparation}.

\bibitem{ivanov08}
B.~A.~Ivanov, R.~S.~Khymyn, and A.~K.~Kolezhuk, 
{\it Pairing of Solitons in Two-Dimensional S=1 Magnets},
Phys.~Rev.~Lett. {\bf 100}, 047203 (2008).


\bibitem{mermin79}
N.~D.~Mermin, 
{\it The topological theory of defects in ordered media},
Rev.~Mod.~Phys. {\bf 51}, 591 (1979).

\bibitem{dadda78}
A.~D'adda, M.~L\"{u}scher, and P.~Di.~Vecchia, 
{\it A $1/n$ Expandable Series of Non-linear $\sigma$ Models with Instantons}, 
Nucl.~Phys.~B {\bf 146}, 63 (1978).


\bibitem{manton04}
N. Manton and P. Sutcliffe, 
{\it Topological Solitons}
(Cambridge University Press, Cambridge, England, 2004).


\bibitem{kirkpatrick83} 
S. Kirkpatrick, C. D. Gelatt Jr., and M. P. Vecchi, 
{\it Optimization by simulated annealing}, 
Science \textbf{220}, 671 (1983).

\bibitem{supplemental} See related discussion in the supplemental materials.


\bibitem{chubukov94}
A. V. Chubukov, S. Sachdev, and T. Senthil, 
\textit{Quantum phase transitions in frustrated two-dimensional antiferromagnets}, 
Nucl. Phys. B \textbf{426}, 601 (1994).

\bibitem{ishida97}
K. Ishida, M. Morishita, K. Yawata, and Hiroshi Fukuyama, 
\textit{Low Temperature Heat-Capacity Anomalies in Two-Dimensional Solid 3He}, 
Phys. Rev. Lett. \textbf{79}, 3451 (1997).

\bibitem{lee08}
P. A. Lee, 
\textit{An End to the Drought of Quantum Spin Liquids}, 
Science \textbf{321}, 1306 (2008).

\bibitem{cheng11}
J. G. Cheng, G. Li, L. Balicas, J. S. Zhou, J. B. Goodenough, 
C. Xu, and H. D. Zhou, 
\textit{High-Pressure Sequence of Ba$_3$NiSb$_2$O$_9$ 
Structural Phases: New S = 1 Quantum Spin Liquids Based on Ni$^{2+}$},
Phys. Rev. Lett. \textbf{107}, 197204 (2011).

\bibitem{xu12prl}
C. Xu, F. Wang, Y. Qi, L. Balents, and M. P. A. Fisher,
\textit{Spin Liquid Phases for Spin-1 Systems on the Triangular Lattice},
Phys. Rev. Lett. \textbf{108}, 087204 (2012).

\bibitem{bieri12prb}
S. Bieri, M. Serbyn, T. Senthil, and P. A. Lee,
\textit{Paired chiral spin liquid with a Fermi surface in $S=1$ model on the triangular lattice},
Phys. Rev. B \textbf{86}, 224409 (2012).

\end{thebibliography}

\begin{thebibliography}{101}

\bibitem{penc11-book.chapter}
K. Penc and A. Lauchli, 
{\it Introduction to frustrated magnetism}, (Springer-Verlag Berlin Heidelberg 2011), chapter~13.

\bibitem{ivanov08}
B.~A.~Ivanov, R.~S.~Khymyn, and A.~K.~Kolezhuk, 
{\it Pairing of Solitons in Two-Dimensional S=1 Magnets},
Phys.~Rev.~Lett. {\bf 100}, 047203 (2008).

\bibitem{Toth12}
T. A. T\'oth, A. M. L\"auchli, F. Mila, and K. Penc,
{\it Competition between two- and three-sublattice ordering for S=1 spins on the square lattice},
Phys.~Rev. B {\bf 85}, 140403(R) (2012).

\bibitem{dadda78}
A.~D'adda, M.~L\"{u}scher, and P.~Di.~Vecchia, 
{\it A $1/n$ Expandable Series of Non-linear $\sigma$ Models with Instantons}, 
Nucl.~Phys.~B {\bf 146}, 63 (1978).

\bibitem{kirkpatrick83} 
S. Kirkpatrick, C. D. Gelatt Jr., and M. P. Vecchi, 
{\it Optimization by simulated annealing}, 
Science \textbf{220}, 671 (1983).


\end{thebibliography}
\end{document}